\documentclass[showpacs,amssymb,preprint,preprintnumbers,nofootinbib,superscriptaddress]{revtex4}
\usepackage{amsmath}
\usepackage{graphicx}
\usepackage{latexsym}
\usepackage{amsfonts}
\usepackage{url,hyperref}
\usepackage{bm}
\usepackage{textcomp}
\usepackage{color}

\newcommand{\beq}{\begin{equation}}
\newcommand{\eeq}{\end{equation}}
\def\bea{\begin{eqnarray}}
\def\eea{\end{eqnarray}}

\begin{document}

\title{Noise kernels of stochastic gravity in conformally-flat spacetimes}
\author{H. T. Cho}
\email[Email: ]{htcho@mail.tku.edu.tw}
\affiliation{Department of Physics, Tamkang University, Tamsui, New Taipei City, Taiwan}
\author{B. L. Hu}
\email[Email: ]{blhu@umd.edu}
\affiliation{Maryland Center for Fundamental Physics, Department of Physics, University of Maryland, College Park, Maryland 20742-4111, USA}

\begin{abstract}
The central object in the theory of semiclassical stochastic gravity is the noise kernel which is the symmetric two point correlation function of the stress-energy tensor. Using the corresponding Wightman functions in Minkowski, Einstein and open Einstein spaces, we construct the noise kernels of a conformally coupled scalar field in these spacetimes. From them we  show that the noise kernels in conformally-flat spacetimes, including the Friedmann-Robertson-Walker universes, can be obtained in closed analytic forms by using a combination of conformal and coordindate transformations.
\end{abstract}

\maketitle

%
%

\section{Introduction}\label{sec:intro}


In semiclassical gravity (SCG) the effect of quantum fields on the background spacetime \cite{BirDav} is accounted for by solving the Einstein equation with the expectation value of the quantum matter field's stress-energy tensor as the source, known as the semiclassical Einstein equation (SCE). To tackle this so-called `backreaction' problem \cite{cosbkrn,bhbkrn} one needs first to find a finite expression of the expectation value of the stress-energy tensor by suitable regularization and renormalization procedures. The resulting expression for the stress-energy tensor is complicated and usually cannot be expressed in a closed analytic form. For example, even for the generic Schwarzschild black hole, it takes considerable work  to get an approximate but numerically accurate expression for the stress-energy tensor \cite{AHS}.

One exception is for the conformally invariant fields, especially in conformally-flat spacetimes. It is possible to derive the stress-energy tensor in closed form by integrating the local expressions of the  corresponding conformal anomaly \cite{ChrFul,BroOtt}. One subtlety in this procedure is in the choice of the vacuum state, even in flat space, be it the Minkowski or the Rindler vacua. This point has been discussed in some detail in the paper by Candelas and Dowker \cite{CanDow} where conformally-flat spacetimes are classified into two categories with the help of the Penrose diagrams. With the Minkowski vacuum there are the spatially flat de Sitter, the flat Friedmann-Robertson-Walker (FRW) universe, the Einstein universe, the global de Sitter, and the closed RW universe. While with the Rindler vacuum one has the open Einstein universe, the Milne universe, the open FRW universe, and the static de Sitter. On the other hand, the two vacua are related by `thermalization' \footnote{This terminology used by \cite{CanDow} refers to a transformation between states, one of which is thermal. It has nothing to do with the thermalization process in nonequilibrium systems.}, in the way that the Minkowski vacuum is  the `thermalization' of the Rindler vacuum. In the next section we shall explore this situation by looking at the Wightman functions in various conformally-flat spacetimes.

Semiclassical gravity based on the expectation value of the stress-energy tensor provides a mean field theory description. To take into account the effects of the matter field quantum fluctuations and the induced metric fluctuations of the spacetime (obtained as a solution of the SCE), one needs to invoke stochastic gravity \cite{HuVer}.  In this theory the matter quantum fluctuations manifest as a stochastic force term in the Einstein equation, resulting in the so-called Einstein-Langevin equation \cite{ELE} -- a physical way of derivation and understanding is by means of  the Brownian motion paradigm (see, e.g., \cite{HPZ} and references therein). The correlator of the stochastic force is given by the noise kernel, which is also the symmetric two point correlation function of the stress-energy tensor. Although the noise kernel is a finite quantity, an intermediate regularization is nevertheless needed. In \cite{PhiHu} a general formula for the noise kernel of a scalar field in terms of higher covariant derivatives of the Green functions is provided.  In this paper we first find the Wightman functions of conformal scalar fields and then use the expressions of \cite{PhiHu} to evaluate the noise kernels  in Minkowski, Einstein, and open Einstein spacetimes.

It is proven in \cite{EBRAH} that the noise kernels of a conformally coupled scalar field in two conformally related spacetimes, $\tilde{g}_{\mu\nu}(x)=\Omega^{2}(x)g_{\mu\nu}(x)$, are related simply by $\tilde{N}_{\mu\nu\alpha'\beta'}(x,x')=\Omega^{-2}(x)N_{\mu\nu\alpha'\beta'}(x,x')\Omega^{-2}(x')$ (here primes on indices denote tensor indices at the point $x'$ and unprimed ones denote indices at the point $x$). Hence, from the noise kernels in Minkowski and Einstein spacetimes, it is possible to obtain the various FRW noise kernels by the appropriate conformal transformations. With additional coordinate transformations, the static de Sitter case can be dealt with. Lastly, the Rindler noise kernel will also be presented. These are the main content in Sections III and IV. Conclusions and discussions are given in Section V.

%
%

\section{Wightman functions in Einstein universes}\label{sec:wightman}

In this section we consider the Wightman functions of a conformally coupled scalar field in Minkowski, closed Einstein and open Einstein spacetimes. Using these functions we can derive the corresponding noise kernels in these spacetimes and also the related FRW spacetimes. We shall do that in the next two sections.

In the Minkowski spacetime, $ds^{2}_{M}=-dt^{2}+dr^{2}+r^{2}(d\theta^2+\sin^{2}\!\theta\ \! d\phi^{2})$, the corresponding Wightman function of a scalar field is just
\begin{eqnarray}
G^{+}_{M}(x,x')=\frac{1}{4\pi^{2}(-\Delta t^{2}+\Delta s^{2})}\label{minkowskiwightman}
\end{eqnarray}
where $\Delta t=t-t'$ and $\Delta s=\sqrt{r^2+r'^2-2rr'(\cos\theta\cos\theta'
+\sin\theta\sin\theta'\cos(\phi-\phi'))}$.

Next, we look at the Einstein universe ($R\times S^{3}$) with the metric
\begin{eqnarray}
ds^{2}_{E}=-dt^{2}_{E}+a^{2}\left(d\chi^{2}+\sin^{2}\!\chi\ \!d\theta^{2}+\sin^{2}\!\chi\ \!\sin^{2}\!\theta\ \! d\phi^{2}\right).
\end{eqnarray}
where $\pi\geq\chi\geq 0$. Under the coordinate transformation,
\begin{eqnarray}
t\pm r=a\tan\left(\frac{t_{E}/a\pm \chi}{2}\right),
\end{eqnarray}
the Einstein universe metric becomes
\begin{eqnarray}
ds^{2}_{E}=4\cos^{2}\left(\frac{t_{E}/a+\chi}{2}\right)\cos^{2}\left(\frac{t_{E}/a
-\chi}{2}\right)ds^{2}_{M}.
\end{eqnarray}
This indicates that the Einstein universe is conformally related to the Minkowski spacetime with the conformal factor
\begin{eqnarray}
\Omega(x)=2\cos\left(\frac{t_{E}/a+\chi}{2}\right)\cos\left(\frac{t_{E}/a
-\chi}{2}\right).
\end{eqnarray}
Similarly, the Wightman function $G_{E}^{+}(x,x')$ of a conformally coupled scalar field in the Einstein universe is related to the corresponding $G_{M}^{+}(x,x')$ by \begin{eqnarray}
G_{E}^{+}(x,x')=\Omega^{-1}(x)G_{M}^{+}(x,x')\Omega^{-1}(x').
\end{eqnarray}
Since the spatial manifold is homogeneous, one can take $\theta=\theta'$ and $\phi=\phi'$ without loss of generality. Then
\begin{eqnarray}
G_{M}^{+}&=&\left(\frac{1}{4\pi^{2}}\right)
\left[a\tan\left(\frac{t_{E}/a+\chi}{2}\right)-a\tan\left(\frac{t'_{E}/a+\chi'}{2}\right)\right]^{-1}\nonumber\\
&&\ \ \ \ \ \ \left[-a\tan\left(\frac{t_{E}/a-\chi}{2}\right)+a\tan\left(\frac{t'_{E}/a-\chi'}{2}\right)\right]^{-1},
\end{eqnarray}
and \cite{BGO}
\begin{eqnarray}
G_{E}^{+}=\frac{1}{8\pi^{2}a^{2}}\left[\cos\left(\frac{\Delta t_{E}}{a}\right)-\cos\left(\frac{\Delta s}{a}\right)\right]^{-1}.\label{closedwightman}
\end{eqnarray}
Here $\Delta s=a\Delta\chi$. In general it is the geodesic distance between two points on the spatial $S^{3}$, $\Delta s=a(\Delta\gamma)$ with
\begin{equation}
\cos\Delta\gamma=\cos\chi\cos\chi'
+\sin\chi\sin\chi'(\cos\theta\cos\theta'+\sin\theta\sin\theta'\cos(\phi-\phi')).
\end{equation}

Finally we come to the open Einstein universe ($R^1\times H^3$) with the metric
\begin{equation}
ds^{2}_{E}=-dt^{2}_{O}+a^{2}\left(d\chi^{2}+\sinh^{2}\!\chi\ \!d\theta^{2}+\sinh^{2}\!\chi\ \!\sin^{2}\!\theta\ \! d\phi^{2}\right).\label{openEinstein}
\end{equation}
where $\chi\geq 0$. Since one could go from the sphere $S^3$ to the hyperboloid $H^3$ by just changing $a\rightarrow ia$, naively one would assume that the Wightman function in this case can be obtained from the Einsten universe one by
\begin{eqnarray}
\left.G^{+}_{E}\right|_{a\rightarrow ia}=-\frac{1}{8\pi^{2}a^{2}}\left[\cosh\left(\frac{\Delta t_{E}}{a}\right)-\cosh\left(\frac{\Delta s}{a}\right)\right]^{-1}.\label{atoia}
\end{eqnarray}
However, this is in fact not the case. Open Einstein universe is a static spacetime and therefore it has a unique vacuum. The corresponding Wightman function was derived exactly by Bunch \cite{Bunch} giving
\begin{eqnarray}
G^{+}_{O}(x,x')=\frac{\Delta s/a}{4\pi^{2}\ \!\sinh(\Delta s/a)(-\Delta t_{O}^{2}+\Delta s^{2})}.\label{openwightman}
\end{eqnarray}
where $\Delta s=a\Delta\gamma$ with
\begin{equation}
\cosh\Delta\gamma=\cosh\chi\cosh\chi'
-\sinh\chi\sinh\chi'(\cos\theta\cos\theta'+\sin\theta\sin\theta'\cos(\phi-\phi')).
\end{equation}
is again the geodesic distance on $H^3$. This Wightman function is not the same as that in Eq.~(\ref{atoia}). Therefore the vacua of the closed and the open Einstein universes are not conformally related even though the spacetimes are.

Along this line the authors in \cite{CanDow} have classified the conformally-flat spacetimes into two classes, one with the conformal vacuum of the Minkowski spacetime and the other with that of the open Einstein spacetime. On the other hand, these two classes are related by `thermalization'. This can be seen using the Wightman functions as follows. From Eq.~(\ref{openwightman}) the Wightman function of a thermal state in the open Einstein universe with temperature $T=1/\beta=1/2\pi a$ can be written as
\begin{eqnarray}
G^{+}_{O}(x,x')_{thermal}=\sum_{n=-\infty}^{\infty}\frac{\Delta s/a}{4\pi^{2}\ \!\sinh(\Delta s/a)[-(\Delta t_{O}+in\beta)^{2}+\Delta s^{2}]}\label{thermal}
\end{eqnarray}
which satisfies the KMS condition. The summation over $n$ can be done using the formula \cite{GraRyz}
\begin{eqnarray}
\sum_{n=-\infty}^{\infty}\frac{1}{(n+A)^{2}+B^{2}}=\frac{\pi\sinh2\pi B}{B(\cosh2\pi B-\cos2\pi A)}.
\end{eqnarray}
Then Eq.~(\ref{thermal}) becomes
\begin{eqnarray}
G^{+}_{O}(x,x')_{thermal}=-\frac{1}{8\pi^{2}a^{2}}\left[\cosh\left(\frac{\Delta t_{O}}{a}\right)-\cosh\left(\frac{\Delta s}{a}\right)\right]^{-1}
\end{eqnarray}
which is the same as Eq.~(\ref{atoia}) since $t_{O}=t_{E}$. This `thermalization' relation between the open and the closed Einstein universes also holds between the static and the global or spatially flat de Sitter, as well as the Rindler and the Minkowski spacetimes.

%
%

\section{Noise kernels related to flat and closed Friedmann-Robertson-Walker spacetimes}\label{sec:closedFRW}

Here we are dealing with conformally invariant scalar fields. Their noise kernels in conformally related spacetimes have the transformation property \cite{EBRAH}
\begin{eqnarray}
\tilde{N}_{\mu\nu\alpha'\beta'}(x,x')
=\Omega(x)^{-2}N_{\mu\nu\alpha'\beta'}(x,x')\Omega(x')^{-2}.
\end{eqnarray}
Since the FRW spacetimes are conformal to the Minkowski and the Einstein universes, we can obtain the noise kernels of various FRW spacetimes from the Einstein universe ones.

In this section we shall concentrate on the flat and the closed universes and leave the discussions on the open FRW ones to the next section. To compute the noise kernels in the Einstein universes, we make use of the formula \cite{PhiHu},
\begin{equation}
 N_{\mu\nu\alpha'\beta'}  = \bar K_{\mu\nu\alpha'\beta'}  + g_{\mu\nu}   \bar K_{\alpha'\beta'} + g_{\alpha'\beta'} \bar K'_{\mu\nu}
 + g_{\mu\nu}g_{\alpha'\beta'} \bar K\label{phillipshu}
\end{equation}
where
\begin{eqnarray}
9  \bar K_{\mu\nu\alpha'\beta'} &=&
%
4\,\left( G{}\!\,_{;}{}_{\alpha'}{}_{\nu}\,G{}\!\,_{;}{}_{\beta'}{}_{\mu} +
    G{}\!\,_{;}{}_{\alpha'}{}_{\mu}\,G{}\!\,_{;}{}_{\beta'}{}_{\nu} \right)
%
+ G{}\!\,_{;}{}_{\alpha'}{}_{\beta'}\,G{}\!\,_{;}{}_{\mu}{}_{\nu} +
  G\,G{}\!\,_{;}{}_{\mu}{}_{\nu}{}_{\alpha'}{}_{\beta'} \cr
%
&& -2\,\left( G{}\!\,_{;}{}_{\nu}\,G{}\!\,_{;}{}_{\alpha'}{}_{\mu}{}_{\beta'} +
    G{}\!\,_{;}{}_{\mu}\,G{}\!\,_{;}{}_{\alpha'}{}_{\nu}{}_{\beta'} +
    G{}\!\,_{;}{}_{\beta'}\,G{}\!\,_{;}{}_{\mu}{}_{\nu}{}_{\alpha'} +
    G{}\!\,_{;}{}_{\alpha'}\,G{}\!\,_{;}{}_{\mu}{}_{\nu}{}_{\beta'} \right)  \cr
%
&& + 2\,\left(
G{}\!\,_{;}{}_{\mu}\,G{}\!\,_{;}{}_{\nu}\,{R{}_{\alpha'}{}_{\beta'}} +
    G{}\!\,_{;}{}_{\alpha'}\,G{}\!\,_{;}{}_{\beta'}\,{R{}_{\mu}{}_{\nu}} \right)  \cr
%
&& - \left( G{}\!\,_{;}{}_{\mu}{}_{\nu}\,{R{}_{\alpha'}{}_{\beta'}} +
  G{}\!\,_{;}{}_{\alpha'}{}_{\beta'}\,{R{}_{\mu}{}_{\nu}}\right) G
%
 +{\frac{1}{2}}  {R{}_{\alpha'}{}_{\beta'}}\,{R{}_{\mu}{}_{\nu}} {G^2}\\
 36  \bar K'_{\mu\nu} &=&
%
8 \left(
 -  G{}\!\,_{;}{}_{\rho'}{}_{\nu}\,G{}\!\,_{;}{}^{\rho'}{}_{\mu}
 + G{}\!\,_{;}{}_{\nu}\,G{}\!\,_{;}{}_{\rho'}{}_{\mu}{}^{\rho'} +
  G{}\!\,_{;}{}_{\mu}\,G{}\!\,_{;}{}_{\rho'}{}_{\nu}{}^{\rho'}
\right)\cr &&
%
4 \left(
    G{}\!\,_{;}{}^{\rho'}\,G{}\!\,_{;}{}_{\mu}{}_{\nu}{}_{\rho'}
  - G{}\!\,_{;}{}_{\rho'}{}^{\rho'}\,G{}\!\,_{;}{}_{\mu}{}_{\nu} -
  G\,G{}\!\,_{;}{}_{\mu}{}_{\nu}{}_{\rho'}{}^{\rho'}
\right) \cr
%
&& - 2\,{R'}\,\left( 2\,G{}\!\,_{;}{}_{\mu}\,G{}\!\,_{;}{}_{\nu} -
    G\,G{}\!\,_{;}{}_{\mu}{}_{\nu} \right)  \cr
%
&&  -2\,\left( G{}\!\,_{;}{}_{\rho'}\,G{}\!\,_{;}{}^{\rho'} -
    2\,G\,G{}\!\,_{;}{}_{\rho'}{}^{\rho'} \right) \,{R{}_{\mu}{}_{\nu}}
%
 - {R'}\,{R{}_{\mu}{}_{\nu}} {G^2}\\
 36 \bar K &=&
2\,G{}\!\,_{;}{}_{\rho'}{}_{\sigma}\,G{}\!\,_{;}{}^{\rho'}{}^{\sigma}
+ 4\,\left( G{}\!\,_{;}{}_{\rho'}{}^{\rho'}\,G{}\!\,_{;}{}_{\sigma}{}^{\sigma} +
    G\,G{}\!\,_{;}{}_{\rho}{}^{\rho}{}_{\sigma'}{}^{\sigma'} \right)  \cr
&& - 4\,\left( G{}\!\,_{;}{}_{\rho}\,G{}\!\,_{;}{}_{\sigma'}{}^{\rho}{}^{\sigma'} +
    G{}\!\,_{;}{}^{\rho'}\,G{}\!\,_{;}{}_{\sigma}{}^{\sigma}{}_{\rho'} \right)  \cr
&& + R\,G{}\!\,_{;}{}_{\rho'}\,G{}\!\,_{;}{}^{\rho'} +
  {R'}\,G{}\!\,_{;}{}_{\rho}\,G{}\!\,^{;}{}^{\rho} \cr
&& - 2\,\left( R\,G{}\!\,_{;}{}_{\rho'}{}^{\rho'} +
{R'}\,G{}\!\,_{;}{}_{\rho}{}^{\rho} \right)
     G
+  {\frac{1}{2}} R\,{R'} {G^2}  \;.
\end{eqnarray}

Note that the superscript $+$ on $G^+$ has been omitted for notational simplicity. $R_{\mu\nu}$ and $R_{\alpha'\beta'}$ are the Ricci tensor evaluated at the points $x$ and $x'$, respectively; $R$ and $R'$ are the scalar field curvature evaluated at $x$ and $x'$.

\subsection{Flat FRW spacetimes}

Here we evaluate the noise kernel of a flat FRW spacetime,
\begin{eqnarray}
ds^{2}=a^{2}(\eta)\left(-d\eta^{2}+dr^{2}+r^{2}d\theta^{2}+r^{2}\sin^{2}\theta d\phi^{2}\right),
\end{eqnarray}
which is conformal to the Minkowski spacetime. First by plugging the Minkowski Wightman function in Eq.~(\ref{minkowskiwightman}) into the formula in Eq.~(\ref{phillipshu}), one can arrive at the Minkowski noise kernel. For the convenience of later presentation, we shall write the noise kernel in the following manner by defining the coefficient functions $C_{ij}$,
\begin{eqnarray}
N_{\eta\eta\eta'\eta'}(x,x')&=&C_{11}\\
N_{\eta\eta\eta' i'}(x,x')&=&C_{21}s_{i'}\\
N_{\eta\eta i'j'}(x,x')&=&C_{31}s_{i'}s_{j'}+C_{32}g_{i'j'}\\
N_{\eta i\eta j'}(x,x')&=&C_{41}s_{i}s_{j'}+C_{42}g_{ij'}\\
N_{\eta ij'k'}(x,x')&=&C_{51}s_{i}s_{j'}s_{k'}+C_{52}s_{i}g_{j'k'}+C_{53}(g_{ij'}s_{k'}+g_{ik'}s_{j'})\\
N_{ijk'l'}(x,x')&=&C_{61}s_{i}s_{j}s_{k'}s_{l'}+C_{62}(g_{ij}s_{k'}s_{l'}+s_{i}s_{j}g_{k'l'})\nonumber\\
&&\ \ +C_{63}(g_{ik'}s_{j}s_{l'}+g_{il'}s_{j}s_{k'}+g_{jk'}s_{i}s_{l'}+g_{jl'}s_{i}s_{k'})\nonumber\\
&&\ \ +C_{64}(g_{ik'}g_{jl'}+g_{il'}g_{jk'})+C_{65}g_{ij}g_{k'l'}\label{noisekernel}
\end{eqnarray}
where $s_{i}=\nabla_{i}(\Delta s)$ and $s_{j'}=\nabla_{j'}(\Delta s)$ are the derivatives on the spatial geodesic distance $\Delta s$ between $x$ and $x'$. Also, $g_{ij'}$ is the parallel transport bivector such that $s_{i}=-g_{i}{}^{j'}s_{j'}$. $C_{ij}$ are functions of $\Delta\eta$ and $\Delta s$. For the Minkowski spacetime we have
\begin{eqnarray}
(C_{11})_{M}&=&\frac{3\Delta s^{4}+10\Delta s^{2}\Delta\eta^{2}+3\Delta\eta^{4}}{12\pi^{4}(-\Delta\eta^{2}+\Delta s^{2})^{6}}\ \ ;
\ \ (C_{21})_{M}=\frac{2\Delta s\Delta\eta(\Delta s^{2}+\Delta\eta^{2})}{3\pi^{4}(-\Delta\eta^{2}+\Delta s^{2})^{6}}\nonumber\\
(C_{31})_{M}&=&\frac{4\Delta\eta^{2}\Delta s^{2}}{3\pi^{4}(-\Delta\eta^{2}+\Delta s^{2})^{6}}\ \ ;
\ \ (C_{32})_{M}=\frac{1}{12\pi^{4}(-\Delta\eta^{2}+\Delta s^{2})^{4}}\nonumber\\
(C_{41})_{M}&=&-\frac{\Delta s^{2}(3\Delta\eta^{2}+\Delta s^{2})}{3\pi^{4}(-\Delta\eta^{2}+\Delta s^{2})^{6}}\ \ ;
\ \ (C_{42})_{M}=-\frac{\Delta\eta^{2}+\Delta s^{2}}{6\pi^{4}(-\Delta\eta^{2}+\Delta s^{2})^{5}}\nonumber\\
(C_{51})_{M}&=&-\frac{4\Delta\eta\ \!\Delta s^{3}}{3\pi^{4}(-\Delta\eta^{2}+\Delta s^{2})^{6}}\ \ ;
\ \ (C_{52})_{M}=0\ \ ;
\ \ (C_{53})_{M}=-\frac{\Delta\eta\Delta s}{3\pi^{4}(-\Delta\eta^{2}+\Delta s^{2})^{5}}\nonumber\\
(C_{61})_{M}&=&\frac{4\Delta s^{4}}{3\pi^{4}(-\Delta\eta^{2}+\Delta s^{2})^{6}}\ \ ;
\ \ (C_{62})_{M}=0\ \ ;
\ \ (C_{63})_{M}=\frac{\Delta s^{2}}{3\pi^{4}(-\Delta\eta^{2}+\Delta s^{2})^{5}}\nonumber\\
(C_{64})_{M}&=&\frac{1}{6\pi^{4}(-\Delta\eta^{2}+\Delta s^{2})^{4}}\ \ ;
\ \ (C_{65})_{M}=-\frac{1}{12\pi^{4}(-\Delta\eta^{2}+\Delta s^{2})^{4}}
\end{eqnarray}
Since the flat FRW spacetime is conformal to the Minkowski one with the conformal factor $\Omega(x)=a(\eta)$, the corresponding coefficients are just given by
\begin{eqnarray}
(C_{ij})_{fFRW}=a^{-2}(\eta)(C_{ij})_{M}a^{-2}(\eta')
\end{eqnarray}
The most prominent example here is the de Sitter spacetime in spatially flat coordinates with $a(\eta)=-1/H\eta$ where $H$ is the Hubble constant. The corresponding vacuum is the Bunch-Davies vacuum \cite{BunDav} which is conformal to the Minkowski one. Many applications in cosmology including those on inflation \cite{Linde} invoke this metric.

This noise kernel has been considered previously in \cite{BCAH} in which it is coordinate transformed into de Sitter static coordinates. According to the `thermalization' relation discussed earlier, the resulting noise kernel is the one in thermal vacuum with temperature $H/2\pi$ with respect to the static vacuum. The behaviors of the noise kernel near the de Sitter horizon is then investigated and is compared with that near the black hole horizon in the Hartle-Hawking thermal vacuum \cite{HarHaw}.

\subsection{Closed FRW spacetimes}

Next, we look at the closed FRW spacetimes with the metric
\begin{eqnarray}
ds^{2}=a^{2}(\eta)\left(-d\eta^{2}+d\chi^{2}+\sin^{2}\!\chi\ \! d\theta^{2}+\sin^{2}\!\chi\ \! \sin^{2}\!\theta\ \! d\phi^{2}\right),
\end{eqnarray}
which is conformal to the Einstein universe with the conformal factor $a(\eta)$. Hence we need to first evaluate the noise kernel of the Einstein universe. This can be done as in the last subsection by plugging the Wightman function in Eq.~(\ref{closedwightman}) into the formula in Eq.~(\ref{phillipshu}). The results are the following.
\begin{eqnarray}
(C_{11})_{E}&=&\frac{4-\cos^{2}\Delta\eta-6\cos\Delta\eta\cos\Delta s-\cos^{2}\Delta s+4\cos^{2}\Delta\eta\cos^{2}\Delta s}{192\pi^{4}(\cos\Delta\eta-\cos\Delta s)^{6}}\nonumber\\
(C_{21})_{E}&=&\frac{\sin\Delta\eta\sin\Delta s(1-\cos\Delta\eta\cos\Delta s)}{48\pi^{4}(\cos\Delta\eta-\cos\Delta s)^{6}}\nonumber\\
(C_{31})_{E}&=&\frac{\sin^{2}\Delta\eta\sin^{2}\Delta s}{48\pi^{4}(\cos\Delta\eta-\cos\Delta s)^{6}}\nonumber\\
(C_{32})_{E}&=&\frac{1}{192\pi^{4}(\cos\Delta\eta-\cos\Delta s)^{4}}\nonumber
\end{eqnarray}
\begin{eqnarray}
(C_{41})_{E}&=&-\frac{(1+\cos\Delta\eta)(1-\cos\Delta s)(2-\cos\Delta\eta+\cos\Delta s-2\cos\Delta\eta\cos\Delta s)}{96\pi^{4}(\cos\Delta\eta-\cos\Delta s)^{6}}\nonumber\\
(C_{42})_{E}&=&-\frac{1-\cos\Delta\eta\cos\Delta s}{96\pi^{4}(\cos\Delta\eta-\cos\Delta s)^{5}}\nonumber\\
(C_{51})_{E}&=&-\frac{\sin\Delta\eta\sin\Delta s(1+\cos\Delta\eta)(1-\cos\Delta s)}{48\pi^{4}(\cos\Delta\eta-\cos\Delta s)^{6}}\ \ ;\ \
(C_{52})_{E}=0\nonumber\\
(C_{53})_{E}&=&-\frac{\sin\Delta\eta\sin\Delta s}{96\pi^{4}(\cos\Delta\eta-\cos\Delta s)^{5}}\nonumber\\
(C_{61})_{E}&=&\frac{(1+\cos\Delta\eta)^{2}(1-\cos\Delta s)^{2}}{48\pi^{4}(\cos\Delta\eta-\cos\Delta s)^{6}}\ \ ;\ \
(C_{62})_{E}=0\ \ ;\ \
(C_{63})_{E}=\frac{(1+\cos\Delta\eta)(1-\cos\Delta s)}{96\pi^{4}(\cos\Delta\eta-\cos\Delta s)^{5}}\nonumber\\
(C_{64})_{E}&=&\frac{1}{96\pi^{4}(\cos\Delta\eta-\cos\Delta s)^{4}}\ \ ;\ \
(C_{65})_{E}=-\frac{1}{192\pi^{4}(\cos\Delta\eta-\cos\Delta s)^{4}}
\end{eqnarray}
Again the coefficients for the closed FRW spacetimes are given by
\begin{eqnarray}
(C_{ij})_{cFRW}=a^{-2}(\eta)(C_{ij})_{E}a^{-2}(\eta')
\end{eqnarray}
For $a(\eta)=\alpha/\sin\eta$, where $\alpha$ is a constant, we have the de Sitter spacetime in the global coordinates .

%
%

\section{Noise kernels related to open FRW spacetimes}\label{sec:closedFRW}

In this section we consider the noise kernels related to the open FRW spacetimes which are conformal to the open Einstein universe. Thus we first work out the noise kernel in the open Einstein universe and then obtain those in the open FRW spacetimes by conformal transformations. On the other hand, we need to make a further coordinate transformation to arrive at the noise kernel in static de Sitter spacetime. Finally, to derive the noise kernel of the Rindler space from the open Einstein, it is necessary to make some combination of conformal and coordinate transformations. This is rather complicated to work through. Hence we would choose another route by working directly with the Rindler Wightman function \cite{Dowker,MorVan} to obtain the noise kernel in the last subsection.

\subsection{Open FRW spacetimes and static de Sitter space}

Consider the open FRW spacetimes with the metric
\begin{eqnarray}
ds^{2}=a^{2}(\eta)(-d\eta^{2}+d\chi^{2}+\sinh^{2}\!\chi\ \! d\theta^{2}+\sinh^{2}\!\chi\ \!\sin^{2}\!\theta\ \! d\phi^{2}),
\end{eqnarray}
which is conformal to the open Einstein universe. Starting with the Wightman function of the open Einstein universe \cite{Bunch}, one can obtain the corresponding noise kernel again using the formula in Eq. (\ref{phillipshu}).
\begin{eqnarray}
(C_{11})_{O}
&=&\frac{G^{2}}{9\Delta s^{2}}\left[1+(3+2\Delta s^{2}){\rm csch}^{2}(\Delta s)-6\Delta s\coth(\Delta s){\rm csch}^{2}(\Delta s)+3\Delta s^{2}{\rm csch}^{4}(\Delta s)\right]\nonumber\\
&&\ \ -\frac{8\pi^{2}G^{3}\sinh(\Delta s)}{9\Delta s^{2}}\left[5\Delta s-6\coth(\Delta s)+12\Delta s\ \!{\rm csch}^{2}(\Delta s)\right.\nonumber\\
&&\ \ \ \ \ \ \ \ \ \ \ \ \ \ \ \ \ \ \ \ \ \ \ \ \ \ \ \ \ \left.-6\Delta s^{2}\coth(\Delta s){\rm csch}^{2}(\Delta s)\right]\nonumber\\
&&\ \ +\frac{64\pi^{4}G^{4}\sinh^{2}(\Delta s)}{3\Delta s^{2}}\left[(3+\Delta s^{2})-2\Delta s\coth(\Delta s)+2\Delta s^{2}{\rm csch}^{2}(\Delta s)\right]\nonumber\\
&&\ \ -\frac{4096\pi^{6}G^{5}\sinh^{3}(\Delta s)}{3\Delta s}+\frac{16384\pi^{8}G^{6}\sinh^{4}(\Delta s)}{3}\nonumber
\end{eqnarray}
\begin{eqnarray}
(C_{21})_{O}
&=&\frac{16\pi^{2}\Delta\eta G^{3}\sinh(\Delta s)}{9\Delta s^{3}}\left[\Delta s-\coth(\Delta s)+2\Delta s\ \!{\rm csch}^{2}(\Delta s)-\Delta s^{2}\coth(\Delta s){\rm csch}^{2}(\Delta s)\right]\nonumber\\
&&\ \ +\frac{64\pi^{4}\Delta\eta G^{4}\sinh^{2}(\Delta s)}{3\Delta s^{3}}\left[(2-\Delta s^{2})-2\Delta s^{2}\ \!{\rm csch}^{2}(\Delta s)\right]\nonumber\\
&&\ \ -\frac{2048\pi^{6}\Delta \eta G^{5}\sinh^{3}(\Delta s)}{3\Delta s^{2}}+\frac{16384\pi^{8}\Delta\eta G^{6}\sinh^{4}(\Delta s)}{3\Delta s}\nonumber
\end{eqnarray}
\begin{eqnarray}
(C_{31})_{O}
&=&\frac{G^{2}}{9\Delta s^{2}}\left[1-(3+4\Delta s^{2})\ \!{\rm csch}^{2}(\Delta s)+6\Delta s\coth(\Delta s)\ \!{\rm csch}^{2}(\Delta s)-3\Delta s^{2}\ \!{\rm csch}^{4}(\Delta s)\right]\nonumber\\
&&\ \ +\frac{8\pi^{2}G^{3}\sinh(\Delta s)}{9\Delta s^{3}}\left[(3-5\Delta s^{2})+9\Delta s\coth(\Delta s)-6\Delta s^{2}\ \!{\rm csch}^{2}(\Delta s)\right.\nonumber\\
&&\ \ \ \ \ \ \ \ \ \ \ \ \ \ \ \ \ \ \ \ \ \ \ \ \ \ \ \ \left.-6\Delta s^{3}\coth(\Delta s)\ \!{\rm csch}^{2}(\Delta s)\right]\nonumber\\
&&\ \ +\frac{64\pi^{4}G^{4}\sinh^{2}(\Delta s)}{3\Delta s}\left[\Delta s-2\coth(\Delta s)+2\Delta s\ \!{\rm csch}^{2}(\Delta s)\right]\nonumber\\
&&\ \ -\frac{4096\pi^{6}G^{5}\sinh^{3}(\Delta s)}{3\Delta s}+\frac{16384\pi^{8}G^{6}\sinh^{4}(\Delta s)}{3}\nonumber
\end{eqnarray}
\begin{eqnarray}
(C_{32})_{O}
&=&\frac{2G^{2}}{9\Delta s^{2}}\left[(1+\Delta s^{2})\ \!{\rm csch}^{2}(\Delta s)-2\Delta s\coth(\Delta s){\rm csch}^{2}(\Delta s)+\Delta s^{2}\ \!{\rm csch}^{4}(\Delta s)\right]\nonumber\\
&&\ \ -\frac{8\pi^{2}G^{3}\sinh(\Delta s)}{9\Delta s^{3}}\left[1+\Delta s\coth(\Delta s)+2\Delta s^{2}\ \!{\rm csch}^{2}(\Delta s)\right.\nonumber\\
&&\ \ \ \ \ \ \ \ \ \ \ \ \ \ \ \ \ \ \ \ \ \ \ \ \ \ \ \ \left.-4\Delta s^{3}\coth(\Delta s){\rm csch}^{2}(\Delta s)\right]\nonumber\\
&&\ \ +\frac{64\pi^{4}G^{4}\sinh^{2}(\Delta s)}{3\Delta s^{2}}\nonumber
\end{eqnarray}
\begin{eqnarray}
(C_{41})_{O}
&=&-\frac{8\pi^{2}G^{3}\sinh(\Delta s)}{9\Delta s^{3}}\left[(2-5\Delta s^{2})+10\Delta s\coth(\Delta s)+7\Delta s\ \!{\rm csch}(\Delta s)\right.\nonumber\\
&&\ \ \ \ \ \ \ \ \ \ \ \ \ \ \ \ \ \ \ \ \ \ \ \ \ \ \ \ \left.-7\Delta s^{2}\coth(\Delta s){\rm csch}(\Delta s)-12\Delta s^{2}\ \!{\rm csch}^{2}(\Delta s)\right]\nonumber\\
&&\ \ -\frac{64\pi^{4}G^{4}\sinh^{2}(\Delta s)}{3\Delta s^{2}}\left[(1+\Delta s^{2})-2\Delta s\coth(\Delta s)-4\Delta s\ \!{\rm csch}(\Delta s)\right.\nonumber\\
&&\ \ \ \ \ \ \ \ \ \ \ \ \ \ \ \ \ \ \ \ \ \ \ \ \ \ \ \ \left.+2\Delta s^{2}\coth(\Delta s){\rm csch}(\Delta s)+3\Delta s^{2}\ \!{\rm csch}^{2}(\Delta s)\right]\nonumber\\
&&\ \ +\frac{1024\pi^{6}G^{5}\sinh^{3}(\Delta s)}{3\Delta s}\left[4-\Delta s\ \!{\rm csch}(\Delta s)\right]
-\frac{16384\pi^{8}G^{6}\sinh^{4}(\Delta s)}{3}\nonumber
\end{eqnarray}
\begin{eqnarray}
(C_{42})_{O}
&=&-\frac{56\pi^{2}G^{3}}{9\Delta s^{2}}\left[1-\Delta s\coth(\Delta s)\right]
+\frac{128\pi^{4}G^{4}\sinh(\Delta s)}{3\Delta s}\left[2-\Delta s\ \!\coth(\Delta s)\right]\nonumber\\
&&\ \ -\frac{1024\pi^{6}G^{5}\sinh^{2}(\Delta s)}{3}\nonumber
\end{eqnarray}
\begin{eqnarray}
(C_{51})_{O}
&=&\frac{8\pi^{2}\Delta\eta G^{3}\sinh(\Delta s)}{9\Delta s^{3}}\left[2\Delta s-5\coth(\Delta s)-4{\rm csch}(\Delta s)+2\Delta s\coth(\Delta s){\rm csch}(\Delta s)\right.\nonumber\\
&&\ \ \ \ \ \ \ \ \ \ \ \ \ \ \ \ \ \ \ \ \ \ \ \left.+\Delta s\ \!{\rm csch}^{2}(\Delta s)+4\Delta s^{2}\coth(\Delta s){\rm csch}^{2}(\Delta s)+2\Delta s^{2}\ \!{\rm csch}^{3}(\Delta s)\right]\nonumber\\
&&\ \ +\frac{64\pi^{4}\Delta\eta G^{4}\sinh^{2}(\Delta s)}{3\Delta s^{3}}\left[(3-\Delta s^{2})+2\Delta s\coth(\Delta s)+4\Delta s\ \!{\rm csch}(\Delta s)\right.\nonumber\\
&&\ \ \ \ \ \ \ \ \ \ \ \ \ \ \ \ \ \ \ \ \ \ \ \ \ \ \ \ \left.-4\Delta s^{2}\coth(\Delta s){\rm csch}(\Delta s)-5\Delta s^{2}\ \!{\rm csch}^{2}(\Delta s)\right]\nonumber\\
&&\ \ +\frac{2048\pi^{6}\Delta\eta G^{5}\sinh^{3}(\Delta s)}{3\Delta s^{2}}\left[1-\Delta s\ \!{\rm csch}(\Delta s)\right]-\frac{16384\pi^{8}\Delta \eta G^{6}\sinh^{4}(\Delta s)}{3\Delta s}\nonumber
\end{eqnarray}
\begin{eqnarray}
(C_{52})_{O}
&=&\frac{8\pi^{2}\Delta \eta G^{3}\sinh(\Delta s)}{9\Delta s^{3}}\left[\coth(\Delta s)+\Delta s\ \!{\rm csch}^{2}(\Delta s)-2\Delta s^{2}\coth(\Delta s){\rm csch}^{2}(\Delta s)\right]\nonumber\\
&&\ \ -\frac{64\pi^{4}\Delta\eta G^{4}\sinh^{2}(\Delta s)}{3\Delta s^{3}}\left[1-\Delta s^{2}\ \!{\rm csch}^{2}(\Delta s)\right]\nonumber
\end{eqnarray}
\begin{eqnarray}
(C_{53})_{O}
&=&-\frac{8\pi^{2}\Delta \eta G^{3}}{9\Delta s^{3}}\left[2-\Delta s\coth(\Delta s)-\Delta s^{2}\ \!{\rm csch}^{2}(\Delta s)\right]\nonumber\\
&&\ \ +\frac{128\pi^{4}\Delta \eta G^{4}\sinh(\Delta s)}{3\Delta s^{2}}\left[1-\Delta s\coth(\Delta s)\right]-\frac{1024\pi^{6}\Delta \eta G^{5}\sinh^{2}(\Delta s)}{3\Delta s}\nonumber
\end{eqnarray}
\begin{eqnarray}
(C_{61})_{O}
&=&\frac{G^{2}}{9\Delta s^{2}}\left[1-16\coth(\Delta s){\rm csch}(\Delta s)-(23-14\Delta s^{2})\Delta s^{2}{\rm csch}^{2}(\Delta s)\right.\nonumber\\
&&\ \ \ \ \ -11\Delta s\coth(\Delta s){\rm csch}^{2}(\Delta s)-16\Delta s\ \!{\rm csch}^{3}(\Delta s)\nonumber\\
&&\ \ \ \ \ \left.+32\Delta s^{2}\coth(\Delta s){\rm csch}^{3}(\Delta s)+34\Delta s^{2}{\rm csch}^{4}(\Delta s)\right]\nonumber\\
&&\ \ +\frac{8\pi^{2}G^{3}\sinh(\Delta s)}{9\Delta s^{3}}\left[(24-5\Delta s^{2})+12\Delta s\coth(\Delta s)+12\Delta s\ \!{\rm csch}(\Delta s)\right.\nonumber\\
&&\ \ \ \ \ \ -28\Delta s^{2}\coth(\Delta s){\rm csch}(\Delta s)-47\Delta s^{2}\ \!{\rm csch}^{2}(\Delta s)\nonumber\\
&&\ \ \ \ \ \ \left.+11\Delta s^{3}\coth(\Delta s){\rm csch}^{2}(\Delta s)+16\Delta s^{3}\ \!{\rm csch}^{3}(\Delta s)\right]\nonumber\\
&&\ \ -\frac{64\pi^{4}G^{4}\sinh^{2}(\Delta s)}{3\Delta s^{2}}\left[(3-\Delta s^{2})+2\Delta s\coth(\Delta s)+16\Delta s\ \!{\rm csch}(\Delta s)\right.\nonumber\\
&&\ \ \ \ \ \ \ \ \left.-8\Delta s^{2}\coth(\Delta s){\rm csch}(\Delta s)-13\Delta s^{2}\ \!{\rm csch}^{2}(\Delta s)\right]\nonumber\\
&&\ \ -\frac{4096\pi^{6}G^{5}\sinh^{3}(\Delta s)}{3\Delta s}\left[1-\Delta s\ \!{\rm csch}(\Delta s)\right]+\frac{16384\pi^{8}G^{6}\sinh^{4}(\Delta s)}{3}\nonumber
\end{eqnarray}
\begin{eqnarray}
(C_{62})_{O}
&=&\frac{G^{2}{\rm csch}^{2}(\Delta s)}{9\Delta s^{2}}\left[2(6+\Delta s^{2})-9\Delta s\coth(\Delta s)-3\Delta s^{2}\ \!{\rm csch}^{2}(\Delta s)\right]\nonumber\\
&&\ \ -\frac{8\pi^{2}G^{3}\sinh(\Delta s)}{9\Delta s^{3}}\left[7+\Delta s\coth(\Delta s)+\Delta s^{2}{\rm csch}^{2}(\Delta s)-9\Delta s^{3}\coth(\Delta s){\rm csch}^{2}(\Delta s)\right]\nonumber\\
&&\ \ +\frac{64\pi^{4}G^{4}\sinh^{2}(\Delta s)}{3\Delta s^{2}}\left[1-\Delta s^{2}\ \!{\rm csch}^{2}(\Delta s)\right]\nonumber
\end{eqnarray}
\begin{eqnarray}
(C_{63})_{O}
&=&-\frac{G^{2}\ \!{\rm csch}(\Delta s)}{9\Delta s^{2}}\left[4\coth(\Delta s)-2(2+3\Delta s^{2}){\rm csch}(\Delta s)+11\Delta s\coth(\Delta s){\rm csch}(\Delta s)\right.\nonumber\\
&&\ \ \ \ \ \ \ \ \ \ \ \ \ \ \ \ \ \ \ \ \ \ \ \left.+4\Delta s\ \!{\rm csch}^{2}(\Delta s)-8\Delta s^{2}\coth(\Delta s){\rm csch}^{2}(\Delta s)-7\Delta s^{2}\ \!{\rm csch}^{3}(\Delta s)\right]\nonumber\\
&&\ \ +\frac{8\pi^{2}G^{3}}{9\Delta s^{2}}\left[3-7\Delta s\coth(\Delta s)-11\Delta s\ \!{\rm csch}(\Delta s)\right.\nonumber\\
&&\ \ \ \ \ \ \ \ \ \ \ \ \ \ \ \ \ \ \ \ \ \ \ \ \ \ \ \ \left.+11\Delta s^{2}\coth(\Delta s){\rm csch}(\Delta s)+4\Delta s^{2}\ \!{\rm csch}^{2}(\Delta s)\right]\nonumber\\
&&\ \ -\frac{128\pi^{4}G^{4}\sinh(\Delta s)}{3\Delta s}\left[2-\Delta s\coth(\Delta s)-\Delta s\ \!{\rm csch}(\Delta s)\right]+\frac{1024\pi^{6}G^{5}\sinh^{2}(\Delta s)}{3}\nonumber
\end{eqnarray}
\begin{eqnarray}
(C_{64})_{O}
&=&\frac{G^{2}\ \!{\rm csch}^{2}(\Delta s)}{9\Delta s^{2}}\left[2(2+3\Delta s^{2})-11\Delta s\coth(\Delta s)+7\Delta s^{2}\ \!{\rm csch}^{2}(\Delta s)\right]\nonumber\\
&&\ \ -\frac{88\pi^{2}G^{3}\ \!{\rm csch}(\Delta s)}{9\Delta s}\left[1-\Delta s\coth(\Delta s)\right]+\frac{128\pi^{4}G^{4}}{3}\nonumber
\end{eqnarray}
\begin{eqnarray}
(C_{65})_{O}
&=&-\frac{G^{2}\ \!{\rm csch}^{2}(\Delta s)}{9\Delta s^{2}}\left[2(3+2\Delta s^{2})-9\Delta s\coth(\Delta s)+3\Delta s^{2}\ \!{\rm csch}^{2}(\Delta s)\right]\nonumber\\
&&\ \ +\frac{8\pi^{2}G^{3}\sinh(\Delta s)}{9\Delta s^{3}}\left[2+7\Delta s^{2}\ \!{\rm csch}^{2}(\Delta s)-9\Delta s^{3}\coth(\Delta s)\ \!{\rm csch}^{2}(\Delta s)\right]\nonumber\\
&&\ \ -\frac{64\pi^{4}G^{4}}{3}\label{opennoise}
\end{eqnarray}
where $G$ is the Wightman function in Eq.~(\ref{openwightman}).

As in the flat and the closed FRW cases, the coefficients for the open FRW spacetime is given by the conformal transformation
\begin{equation}
(C_{ij})_{oFRW}=a^{-2}(\eta)(C_{ij})_{O}\ \!a^{-2}(\eta').
\end{equation}
For $a(\eta)=\alpha e^{\eta}$, where $\alpha$ is a constant, we have the Milne universe \cite{GriPod}.

Next, we look at the noise kernel in static de Sitter spacetime. Going from the open Einstein metric, it is necessary to make one conformal and another coordinate transformations to arrive at this metric. From the metric in Eq.~(\ref{openEinstein}), we perform a conformal trasformation with the conformal factor $\Omega=1/\cosh\chi$, that is,
\begin{equation}
ds^{2}=\left(\frac{1}{\cosh^{2}\!\chi}\right)(-d\eta^{2}+d\chi^{2}+\sinh^{2}\!\chi\ \! d\theta^{2}+\sinh^{2}\!\chi\ \!\sin^{2}\!\theta\ \! d\phi^{2})
\end{equation}
Then making a coordinate transformation, $r=\tanh\chi$, we arrive at the static de Sitter metric
\begin{equation}
ds^{2}=-(1-r^{2})d\eta^{2}+(1-r^{2})^{-1}dr^{2}+r^{2}(d\theta^{2}+\sin^{2}\!\theta\ \! d\phi^{2})
\end{equation}
with the horizon at $r=1$.

Hence, to obtain the noise kernel in static de Sitter we need to make the corresponding conformal and coordinate transformations on the open Einstein space noise kernel in Eqs.~(\ref{noisekernel}) and (\ref{opennoise}). The coefficient functions $(C_{ij})_{sdS}$ of the noise kernel in static de Sitter spacetime are given by
\begin{equation}
(C_{ij})_{sdS}=(1-r^{2})^{-1}(C_{ij})_{O}(1-r'^{2})^{-1},
\end{equation}
where the geodesic distance
\begin{eqnarray}
\Delta s=\cosh^{-1}\left\{(1-r^{2})^{-1/2}(1-r'^{2})^{-1/2}\left[1
-rr'(\cos\theta\cos\theta'+\sin\theta\sin\theta'\cos(\phi-\phi'))\right]\right\}.\nonumber\\
\end{eqnarray}
Note that this is the de Sitter noise kernel in the static vacuum which is conformal to the open Einstein or the Rindler vacuum. Due to `thermalization' relation we discussed earlier, the noise kernels in spatially flat de Sitter \cite{BCAH} or the global de Sitter cases are actually in the thermal static vacuum.

\subsection{Rindler space}

In this subsection we would like to consider the noise kernel in the Rindler spacetime. The Rindler space is flat and its vacuum is conformal to the one in open Einstein universe. Therefore, in principle using a combination of conformal and coordinate transformations, as we have done above, it is possible to obtain the noise kernel in Rindler spacetime from that in open Einstein universe. However, the procedure is quite complicated. To see this we look at the metric. With the coordinate transformation,
\begin{eqnarray}
t\pm r=\tanh\left(\frac{\eta\pm \chi}{2}\right),
\end{eqnarray}
the open Einstein universe metric becomes
\begin{eqnarray}
ds^{2}&=&-d\eta^{2}+d\chi^{2}+\sinh^{2}\!\chi\ \! d\theta^{2}+\sinh^{2}\!\chi\ \!\sin^{2}\!\theta\ \! d\phi^{2}\nonumber\\
&=&4[1-(t-r)^{2}]^{-1}[1-(t+r)^{2}]^{-1}(-dt^{2}+dr^{2}+r^{2}d\theta^{2}+r^{2}\sin^{2}\!\theta\ \! d\phi^{2}).
\end{eqnarray}
Thus with a further conformal transformation one can arrive at the Minkowski metric. Going to Cartesian coordinates and then making another coordinate transformation,
\begin{equation}
t=\xi\sinh\tau\ \ \ ;\ \ \ x=\xi\cosh\tau
\end{equation}
we can get the Rindler metric
\begin{equation}
ds^{2}=-\xi^{2}d\tau^{2}+d\xi^{2}+dy^{2}+dz^{2}
\end{equation}

It is nevertheless quite a complicated procedure to implement this series of transformations on the open Einstein noise kernel to obtain the Rindler one. To avoid this we have chosen to work on the Wightman function in Rindler spacetime directly. This Wightman function has been given in \cite{Dowker,MorVan},
\begin{equation}
G^{+}_{R}=\frac{1}{4\pi^{2}}\left(\frac{\alpha}{\xi\xi'\sinh\alpha}\right)\left(\frac{1}{-(\tau-\tau')^{2}+\alpha^{2}}\right),
\end{equation}
where
\begin{equation}
\cosh\alpha=\frac{\xi^{2}+\xi'^{2}+(y-y')^{2}+(z-z')^{2}}{2\xi\xi'}.
\end{equation}
Plugging this into Eq.~(\ref{phillipshu}) one can obtain the noise kernel components in the Rindler spacetime. However, the expressions for these components are much lengthier than the ones in FRW cases so we just display one of them.
\begin{eqnarray}
N_{\tau\tau\tau'\tau'}
&=&\frac{G^{2}}{9\alpha^{2}\xi^{3}\xi'^{3}}\left\{\xi^{3}\xi'^{3}+(1-\xi^{2}\xi'^{2})(\xi^{2}+\xi'^{2})\coth\alpha\ \!{\rm csch}\alpha\right.\nonumber\\
&&\ \ \ \ \ \ \ \ \ \ +\xi\xi'\left[(1+5\alpha^{2})+(2-3\alpha^{2})\xi^{2}\xi'^{2}-2\alpha(4-\xi^{2}\xi'^{2})\coth\alpha\right]{\rm csch}^{2}\alpha\nonumber\\
&&\ \ \ \ \ \ \ \ \ \ +\alpha(1-\xi^{2}\xi'^{2})(\xi^{2}+\xi'^{2})(1-2\alpha\coth\alpha){\rm csch}^{3}\alpha\nonumber\\
&&\ \ \ \ \ \ \ \ \ \ \left.+\alpha^{2}\xi\xi'(7-4\xi^{2}\xi'^{2})\ \!{\rm csch}^{4}\right\}\nonumber\\
&&-\frac{8\pi^{2}G^{3}\sinh\alpha}{9\alpha^{3}\xi^{2}\xi'^{2}}\left\{\xi^{3}\xi'^{3}\left[(2+5\alpha^{2})-6\alpha\coth\alpha\right]\right.\nonumber\\
&&\ \ \ \ \ \ \ \ \ \ -\alpha(\xi^{2}+\xi'^{2})(3-\xi^{2}\xi'^{2}-2\alpha\coth\alpha){\rm csch}\alpha\nonumber\\
&&\ \ \ \ \ \ \ \ \ \ +2\alpha^{2}\xi\xi'\left[4+\xi^{2}\xi'^{2}-\alpha(4-\xi^{2}\xi'^{2})\coth\alpha\right]{\rm csch}^{2}\alpha\nonumber\\
&&\ \ \ \ \ \ \ \ \ \ \left.+\alpha^{3}(1-\xi^{2}\xi'^{2})(\xi^{2}+\xi'^{2}){\rm csch}^{3}\alpha\right\}\nonumber\\
&&+\frac{64\pi^{4}G^{4}\sinh^{2}\alpha}{9\alpha^{2}\xi\xi'}\left\{3\xi\xi'\left[3+(3+\alpha^{2})\xi^{2}\xi'^{2}
-2\alpha\xi^{2}\xi'^{2}\coth\alpha\right]\right.\nonumber\\
&&\ \ \ \ \ \ \ \ \ \ \left.-\alpha(\xi^{2}+\xi'^{2})(5-\alpha\coth\alpha){\rm csch}\alpha+5\alpha^{2}\xi\xi'\ \!{\rm csch}^{2}\right\}\nonumber\\
&&-\frac{1024\pi^{6}G^{5}\sinh^{3}\alpha}{9\alpha}\left\{6\xi\xi'(1+2\xi^{2}\xi'^{2})-\alpha(\xi^{2}+\xi'^{2}){\rm csch}\alpha\right\}\nonumber\\
&&+\frac{16384\pi^{8}G^{6}\sinh^{4}\alpha}{9}\left\{\xi^{2}\xi'^{2}(1+3\xi^{2}\xi'^{2})\right\}
\end{eqnarray}
The other components of the Rindler noise kernel can be derived analogously from Eq.~(\ref{phillipshu}). This completes our consideration on the noise kernels of a conformal scalar field in conformally-flat spacetimes.

%
%

\section{Conclusions and Discussions}\label{sec:conclusion}

In this paper we have considered the noise kernels of a conformal scalar field in conformally-flat spacetimes. First, conformally-flat spacetimes can be classified into two main categories according to the conformal vacuum they admit, namely, the Minkowski or the Rindler vacuum. We explicated this point with the help of the corresponding Wightman functions, from which we derive the noise kernels in three static spacetimes, namely, the Minkowski, Einstein, and open Einstein spacetimes. Basically the conformal transformations between noise kernels in conformally-related spacetimes are quite simple. Therefore, we are able to obtain the noise kernels of a conformal scalar field in flat, closed, and open FRW spacetimes with arbitrary time-dependent scale factors in closed analytic form from the ones in Minkowski, Einstein and open Einstein spacetimes, respectively.

The noise kernel is the two point correlation function of the stress-energy tensor. It represents the backreaction of the quantum fluctuations of the matter field onto the background spacetime obtained as a solution of the semiclassical Einstein equation. Therefore, it is interesting to investigate the behaviors of the noise kernel near horizons as well as initial singularities of various FRW spacetimes in this paper. In \cite{BCAH} the behaviors of the noise kernel near the de Sitter horizon was studied and then compared with that in the Schwarzschild spacetime \cite{EBRAH}. Similar study can be carried out for the noise kernels found here.

In the theory of stochastic gravity the noise kernel characterizes the correlations of the stochastic force. This stochastic force represents the matter field quantum fluctuations whose effect on the background spacetime is obtained by solving the Einstein-Langevin equation.   For example, the correlation of the quantized gravitational perturbations can be found in the graviton noise kernel which is used in the stochastic gravity approach to structure formation \cite{RouVerStrFor}.  Note the Einstein-Langevin equation also contains dissipative terms in balance of the stochastic force. For the FRW cases we are considering, one can start with the static Minkowski, Einstein, and open Einstein cases to evaluate the corresponding influence functionals. Then by conformal transformation \cite{BroOtt} on these influence functionals one can obtain the Einstein-Langevin equations in FRW spacetimes. We are currently working in this direction. We also note related recent work of noise kernel for inhomogeneous spacetimes \cite{SLB} and Riemann curvature correlators \cite{FrobRouVer}.

%
%

\acknowledgments

HTC is supported in part by the National Science Council,
Taiwan, ROC under the Grants No.~NSC102-2112-M-032-002-MY3, and by the National Center for
Theoretical Sciences (NCTS). HTC would like to thank the hospitality of Prof.~Kin-Wang Ng, and BLH that of Prof.~Hsiang-nan Li of the Theory Group of the Institute of Physics at the Academia Sinica, Republic of China, where part of this work was done.

%
%

\end{document}